**Brain perfusion mediates the relationship between miRNA levels and postural control**

Yufen Chen [1,16], Amy A Herrold [2,3,17], Zoran Martinovich [4,17], Anne J Blood [5-7,17], Nicole Vike [8,9,17], Alexa E Walter [10,17], Jaroslaw Harezlak [11,17], Peter H Seidenberg [12], Manish Bhomia [13], Barbara Knollmann-Ritschel [13], James L Reilly [3], Eric A Nauman [8,9,14,18], Thomas M Talavage [8,9,18], Linda Papa [15,18], Semyon Slobounov [10,18,*], Hans C Breiter[3,6,16,*] for the Concussion Neuroimaging Consortium

[1]Center for Translational Imaging, Department of Radiology, Feinberg School of Medicine, Northwestern University, Chicago IL 60611

[2]Edward Hines Jr., VA Hospital, Research Service, Hines, IL 60141

[3]Warren Wright Adolescent Center, Department of Psychiatry and Behavioral Sciences, Feinberg School of Medicine, Northwestern University, Chicago IL 60611

[4]Mental Health Services and Policy Program, Department of Psychiatry and Behavioral Sciences, Northwestern University Feinberg School of Medicine, Chicago, IL 60611

[5]Mood and Motor Control Laboratory, Departments of Neurology and Psychiatry, Massachusetts General Hospital and Harvard Medical School, Boston, MA 02129

[6]Laboratory of Neuroimaging and Genetics, Department of Psychiatry, Massachusetts General Hospital and Harvard Medical School, Boston, MA 02129

[7]Martinos Center for Biomedical Imaging, Department of Radiology, Massachusetts General Hospital and Harvard Medical School, Boston, MA 02129

[8]School of Electrical and Computer Engineering, Purdue University, West Lafayette, IN 47907

[9]Weldon School of Biomedical Engineering, Purdue University, West Lafayette, IN 47907

[10]Department of Kinesiology, Pennsylvania State University, University Park, PA 16802

[11]Department of Epidemiology and Biostatistics, Indiana University, Bloomington, IN 47405

[12]Departments of Orthopaedics & Rehabilitation and Family & Community Medicine, Penn State College of Medicine, Hershey PA 17033

[13]Department of Pathology, Uniformed Services University of the Health Sciences, Bethesda, MD

[14]Department of Basic Medical Sciences, Purdue University, West Lafayette, IN 47907

[15]Department of Emergency Medicine, Orlando Regional Medical Center, Orlando, Florida, USA.

[16]These authors contributed equally as first authors.

[17]These authors contributed equally as second authors.

[18]These authors contributed equally as senior authors.

* Correspondence

* Corresponding Authors:
For project design, management and data collection: Semyon Slobounov (sms18@psu.edu)
For hypotheses and conceptual framework, data analysis and paper development: Hans Breiter (h-breiter@northwestern.edu)

Word count: 6726

**Summary** (150 words)

Transcriptomics, regional cerebral blood flow (rCBF), and a spatial motor virtual reality task were integrated using mediation analysis in a novel demonstration of "imaging omics". Data collected in NCAA Division I football athletes cleared for play before in-season training showed significant relationships in a) elevated levels of miR-30d and miR-92a to elevated putamen rCBF, (b) elevated putamen rCBF to compromised balance scores, and (c) compromised balance scores to elevated miRNA levels. rCBF acted as a mediator variable (minimum 70% mediation, significant Sobel's test) between abnormal miRNA levels and compromised balance scores. Given the involvement of these miRNAs in inflammation and immune function, and that vascular perfusion is a component of the inflammatory response, these findings support a chronic inflammatory model of repetitive head acceleration events (HAEs). rCBF, a systems biology measure, was necessary for miRNA to affect behavior. These results suggest miRNA as a potential diagnostic biomarker for repetitive HAEs.

## Introduction

Athletes who participate in contact/collision sports are exposed to repeated head acceleration events (HAEs). HAEs can have long-term effects, regardless of etiology (*i.e.*, athletics-based versus combat-based) [1]. Three theories about the mechanism of injury from HAEs and concussion have been proposed: *(1) neurovascular decoupling* [2-6], *(2) neuro-inflammation* [7-14], and *(3) diffuse axonal injury* [15-22]. Although these three models are independently supported by strong evidence, vascular function is a fundamental part of the initiation phase of inflammatory response, and its longer-term resolution phase [23, 24]. Thus, two of the three models have some overlap.

The neuro-inflammation model was the focus of a recent study on micro-RNA (miRNA) levels before and over the course of the football season [25] for miRNAs related to inflammatory responses, such as miR-20a, miR-30d, and miR-92a (see also [26-28]). The panel of miRNAs studied had previously been shown to be elevated in subjects visiting emergency rooms for mild to severe traumatic brain injury (TBI), and were correlated with abnormal clinical readings on CT scans [29]. MiRNAs are small (19-28nt) endogenous RNA molecules that modulate gene expression by suppressing targeted messenger RNAs (mRNAs) [30]; they represent a dynamic measure of gene function and are part of the transcriptome [31]. Since each miRNA can target hundreds of mRNAs, miRNAs are involved in a wide variety of cellular processes, including many of those that occur after the initial physical impact in head injuries, such as the initiation of inflammation, or its longer-term resolution [32, 33]. In football players on a single NCAA Division I collegiate team, Papa and colleagues (2018) found nine miRNAs to be significantly elevated at the beginning of the season, relative to matched controls: miR-20a, miR-505*, miR-362-3p, miR-30d, miR-92a, miR-486, miR-195, miR-9-3p and miR-151-5p. The level of these nine miRNAs exhibited a negative relationship to Standard Assessment of Concussion (SAC) scores [25]. Specifically, higher miRNA levels were associated with greater evidence of clinical symptoms (i.e. lower SAC scores) prior to the football season. These abnormalities occurred in players who had not had any head impacts in 16 weeks, and were medically cleared to begin the upcoming season. In the context that the average history of competitive football play in this cohort was 11 years, these observations raised the hypothesis that the elevated levels were residual circulating miRNAs from prior HAEs and concussions, potentially reflecting chronic inflammation.

*In vivo*, there are potentially two ways to develop support for a hypothesis regarding chronic inflammation in athletes participating in collision sports. One is to use positron emission tomography with a ligand related to inflammatory function (e.g., [34]). A second is to determine if there are functional or behavioral implications for this baseline miRNA abnormality in behaviors that are susceptible to HAEs/concussion, and to localize these behavioral alterations with magnetic resonance imaging (MRI) to brain regions whose activity or perfusion is known to contribute to the behavioral process. For this second approach, use of regional perfusion MRI would provide further support for the hypothesis of a residual inflammatory process if elevated miRNA's were connected both (1) to elevated regional brain perfusion and (2) to impaired behavior (e.g., [35]) for functions that are known to be mediated by the affected brain regions. For this study, we used the second approach, and focused on behavioral measures of motor control, using a validated virtual reality (VR) task grounded in the work of Luria [36-39].

The VR task produced computational behavior measures of (i) spatial navigation accuracy ("Spatial Memory"), (ii) sensory-motor reactivity and efficiency of visual-spatial processing ("Reaction Time"), and (iii) postural stability during equilibrium changes ("Balance"), along with (iv) an integrative metric of all three measures together ("Comprehensive" score). Alexander Luria observed that head impacts without clinical signs of trauma may be associated with chronic neurocognitive deficits in (i) spatial orientation and accuracy of navigation, (ii) processing of visual-spatial information (sensory-motor reactivity), or (iii) related coordination functions such as whole body postural control or balance [40-42].

He proposed that these deficits were due to patients' inability to create a *cognitive map* of the perceptual-working space, making them prone to relying on "short-term memory – counting turns". He also noted patients could have problems reorienting themselves to generalized changes in the space around them, and could have problems with balance [36]. Although a number of computer-based neuropsychological tests exist to assess aspects of Luria's triad [i.e., (i) – (iii) above], VR can detect residual abnormalities in the absence of self-reported symptoms by HAE recipients [25, 37-39].

Connecting such motor control abnormalities to regional perfusion in brain regions mediating motor control, such as the basal ganglia, primary sensorimotor cortex, premotor cortex, supplementary motor area, or cerebellum (e.g., [43-47]) would support the specificity of such observations. Numerous studies have reported changes in brain imaging after clinically diagnosed concussion or HAEs without symptoms, as well as in clinically asymptomatic athletes cleared for sports participation [35, 48-55]. These changes include alterations in cerebrovascular integrity, blood flow, and resting-state functional connectivity as measured by various forms of MRI (e.g. [48, 49, 56]). If this relationship encoded a statistical mediation between these three variables, from three distinct spatiotemporal scales of function (Braeutigam, 2018), it would represent a novel control framework between miRNA, regional brain perfusion, and the computational measure of motor control. To date, no such relationships between miRNA levels, regional brain imaging, and computational behavior - what might be called "imaging omics" - have been reported in humans.

This study developed three aims to test for a relationship between miRNAs, regional neuroimaging measures, and behavior, thus supporting a model that miRNA elevations observed before seasonal activities reflect residual elevations of inflammatory miRNAs related to abnormal regional perfusion in brain regions that produce behaviors impacted by previous HAEs. We used a candidate miRNA approach, focusing our analysis on the panel of aforementioned nine miRNAs reported to be elevated in preseason football players by Papa et al 2018 [25]. *First*, we investigated if a three-way set of associations existed between (i) miRNA level, (ii) regional CBF (rCBF) in motor control regions, and (iii) computational behavior of motor control, that did not depend on an intermediate phenotype like rCBF to connect miRNA levels to behavior. A whole-brain, voxelwise statistical approach was used for this exploratory step to determine regions where perfusion was related to either miRNA level or behavior. MiRNAs are a dynamic measure of systems function, and their relationship to behavior may go beyond the idea of an intermediate phenotype in "imaging genetics" [57], wherein brain imaging acts as a common, overlapping node between two associations. *Second*, we sought to place these associations closer to a mechanistic framework using a directed mediation analysis, wherein miRNA levels were always the input variable in the analysis between these three layers of spatiotemporal organization, and behavior was always the output variable. In such a framework, brain imaging is a required variable for the effect of miRNA on behavior, and provides a proof-of-concept that the systems biology (measured by brain imaging) is needed for the effect of miRNA on behavior. Directed mediation requires a specified precedence of each variable relative to the others, wherein flipping the order of the variables in the association shows there is no longer a significant mediation effect. Directed mediation thus requires confirmation of the absence of confounds such as miRNA levels mediating the relationship between rCBF and behavior, and in such context has been argued to be mechanistic [58, 59] as opposed to purely associative [60]. *Third*, across any relationships of miRNA, rCBF and behavior, we sought to determine if there were consistent patterns across all (rather than some or most of) the observed relationships. For illustration purposes, such consistent patterns might include a positive relationship between miRNA level and rCBF as one might expect with inflammatory processes. Other consistent patterns might include a negative relationship between rCBF and behavior, in line with other published results of football players (e.g., [35]) or a negative relationship between miRNA level and behavior, parallel to what Papa and colleagues ([25]) saw with miRNA and SAC scores. Given the potential for false positives with measures from three distinct spatiotemporal scales, we used several methods to insure rigor including corrections for multiple comparisons at each stage of analysis and graphical inspection for quality assurance of associations, along with a formal Cook's Distance analysis of data point influence on results.

## Materials and Methods

### *Subjects*

A total of 24 collegiate male football athletes were recruited for this study. Written informed consent was obtained from all subjects, in accordance with the guidelines of the Pennsylvania State University's Institutional Review Board. Demographics including age, years of play, football team position, and number of previous concussions diagnosed by medical personnel were collected based on self-report and confirmed by the team physician. Preseason MRI scans were collected within one week before the athletic season started. Only 23 subjects received the pre-season MRI scan. Additional subjects were excluded due to missing microRNA or virtual reality scores. For the final analysis, 20 subjects (age=21±2 years) were included for the pre-season MRI. The average number of years of participation in football was 11±4 years and 6 were speed players and 14 were nonspeed players based on classification by [122]. Two of the subjects had two prior concussions, five had one prior concussion and the rest had no prior concussions.

#### *Serum extraction*

Blood samples were collected within a week before the season began. Five mL of blood was placed in a serum separator tube and centrifuged after clotting at room temperature. The serum was placed bar-coded aliquot containers and stored at -70°C until transport to a central laboratory for batch analysis. Personnel performing the laboratory analysis were blinded to the clinical data.

#### *Virtual reality testing*

During each visit, athletes also completed a previously validated virtual reality neurocognitive testing with a 3D TV system (HeadRehab.com) and a head mounted accelerometer. The test included three modules: spatial memory, balance and whole body reaction time. Scores from these modules were combined to generate a comprehensive score.

Following the behavioral neurology of spatial memory problems in veterans with head injury described by Luria [40], the spatial navigation task involved three modules. The first module, a memory test, showed subjects a randomized virtual pathway with multiple turns to a door along with the return trip. Subjects were instructed to repeat the pathway using a joystick, and their accuracy was assessed via correct responses vs. errors. In the balance task, subjects were instructed to hold a tandem Romberg position for all trials. The virtual room was completely still for the first trial for a baseline measure. In the subsequent six trials, the virtual room moved in various directions, and individual alignment with the virtual room was quantified via a pressure plate on which they stood. For the reaction time module, subjects stood feet shoulder width apart with hands on their hips. They were instructed to move their body in the same direction as the virtual room's movements, and the pressure plate measured response time latency. Raw data was run through mathematical algorithms to output scores on a scale of 0 (worst) to 10 (best). In additional to individual component scores, an overall "comprehensive score" was calculated by combining the three test scores [38]. These VR assessments were assessed at the same time as neuroimaging and serum extraction. Detailed descriptions of the modules and their sensitivity and specificity have been described previously [37-39].

*MRI acquisition*

Imaging data were collected on a 3.0T whole body Siemens Prisma scanner (Erlangen, Germany), equipped with a 32-channel head coil. For both pre- and post-season visits, high resolution T1-weighted anatomical images and 3D-background suppressed (BS)-PASL images were acquired with the following parameters: 3D-MPRAGE – TR/TE/TI = 1700/1.77/850 ms, flip angle = 9°, matrix size = 320× 260 ×176, voxel size = 1 mm isotropic, and parallel acceleration factor = 2 (total duration=3min 31s).; 3D-BS-pCASL – PICORE/Q2TIPS labeling scheme, TR/TE = 4600/15.62 ms, matrix size = 128x128x120, voxel size = 1.5mm x 1.5mm x 3mm, 40 axial slices, $TI_1/TI_2$ = 700/1990 ms, parallel acceleration factor = 2, 6 control/tag pairs and 1 M0 acquisition (total duration=6min 2s).

*Cerebral Blood Flow Mapping*

Imaging data were processed using in-house scripts written in Matlab R2016a (Mathworks, Natick, MA) with Statistical Parametric Mapping SPM8 (Wellcome Department of Imaging Neuroscience, London, UK). All ASL data were motion-corrected with the first image of the series as the reference and then co-registered to the high resolution anatomical image. Perfusion weighted images were generated by pairwise subtraction between control and tag images and averaged over the entire time-series. Images were converted to quantitative CBF units in ml/100g/min using the single-blood-compartment model [123]:

$$f = \frac{\lambda \cdot \Delta M}{2\alpha \cdot M_0 \cdot TI_1 \cdot e^{-TI_2/T_{1b}}}$$

Where $f$ represents CBF in quantitative units, ΔM is the perfusion weighted signal, λ is the blood/water tissue partition coefficient (assumed to be 0.9 g/ml [124]), α is the inversion efficiency assumed to be 0.98*0.75=0.735 [125], $M_0$ is the equilibrium magnetization from the $M_0$ acquisition, and $T_{1b}$ is blood T1 assumed to be 1664ms [126]. The quantitative CBF maps were then transformed to MNI template space and upsampled to 1.5mm isotropic resolution based on transformation matrix calculated from the high resolution anatomical image using VBM8 [127]. A three-dimensional 3mm smoothing kernel was applied to the spatially normalized CBF maps to minimize normalization discrepancies.

*miRNA analysis*

RNA isolation Serum samples from athletes were used to isolate total RNA. RNA was isolated using 100 lL of serum using a serum/plasma miRNA isolation kit (Qiagen Inc.) as per manufacturer's recommended protocol. The RNA was eluted in 20 lL of DNAse RNAse free water and stored at -80C for further use. For a quality check of the RNA, a bioanalyzer assay using a small RNA assay was performed to confirm the quality of the RNA.

Droplet digital PCR (ddPCR) For absolute quantitation of miRNAs, we used a ddPCR platform (Bio Rad Inc.). For ddPCR reaction, 10 ng of total RNA was reverse transcribed using the specific miRNA TaqMan assays (Thermofisher Scientific Inc.) as per recommended protocol in a 15 µL total reaction volume; 5 µL of reverse transcribed product was used to set up the real-time PCR reaction using miRNA TaqMan assays; and 20 µL of the final realtime PCR reaction was mixed with 70 µL of droplet oil in a droplet generator (Bio Rad Inc.). Following the droplet formation, the PCR reaction was performed as per recommended thermal cycling conditions. The final PCR product within the droplets was analyzed in a droplet reader (Bio-Rad Inc). The total positive and negative droplets were measured and the concentration of the specific miRNA/μL of the PCR reaction was determined. All the reactions were performed in duplicates.

*Statistical Analysis*

CBF and VR scores correlation Statistical Parametric Mapping (SPM8, Wellcome Trust Centre for Neuroimaging, UK) was used to calculate voxelwise statistics on the preseason CBF maps. Each of the VR scores was entered as a covariate in a one sampled t-test, and T-contrasts for positive and negative correlations between preseason CBF and VR score were used to locate areas with significant correlations. Residual maps were input into AFNI 3dFWHMx with spatial autocorrelation function option to estimate the smoothness. This was then used as input in 3dClustSim to calculate the cluster size for a cluster-corrected threshold of p=0.05, using a cluster-forming p-value of 0.005. The cluster threshold for this analysis was 203 voxels. Anatomical localization of all clusters were determined based on the Harvard-Oxford cortical (48 regions) and subcortical (21 regions) atlases [128] and a probabilistic cerebellar atlas with 28 anatomical regions [129]. The percentages listed in tables represent the percentage of each anatomical region occupied by the cluster.

CBF and miRNA levels correlation A similar approach as described above was used to determine the correlation between preseason CBF and each of the preseason miRNA levels. The cluster threshold for each miRNA analysis to reach cluster corrected p=0.05 is listed in Table 2.

Influence Testing (Cook's Distance computation) Before data was submitted for three way association (below), using the outcomes of the CBF and VR analysis and the CBF and miRNA levels analysis, we performed a influence assessment to rule out associations being driven by a minority of the data. Using the program STATA, the Cook's Distance method was used to determine significant outlier effects on regression analyses. First, miRNA values were regressed against MRI data (maxvox and cluster). Any Cook's values (D) greater than five signifies a likely significant outlier effect on regression results. Regression analyses of miRNA 505 and 486 against mean cluster rCBF and maxvox exhibited $D > 5$ (mi505 – cluster 1 cluster/maxvox=5.7/1.54, cluster2 cluster/maxvox=5.94/2.48; mi486 – cluster 1 cluster/maxvox=14.8/13.9, cluster2 cluster/maxvox=16.5/12.4). These outliers were removed from the data set and regressions were rerun. After removal of these outliers, the regressions were no longer significant. Balance data were also regressed against miRNA, maxvox, and cluster; no regressions resulted in $D > 5$.

CBF, miRNA and VR scores interaction This analysis evaluated three-way associations across these three measures. To pinpoint regions where there were significant correlations between CBF and miRNA and CBF and VR scores, the CBF and VR score correlation results were used as a mask for small volume correction in the CBF and miRNA correlation analysis. This approach performs multiple comparisons correction only within voxels that had a significant CBF and miRNA correlation. Clusters that survive familywise error correction at $p<0.05$ were reported. To test the robustness of our results, we also reversed the direction of the analysis by using the CBF and miRNA results as mask for the CBF and VR scores correlations. The same clusters were identified with slightly different T values and location of voxel with maximum T value. Therefore, only the VR score masked CBF and miRNA correlation results were reported.

Mediation Analysis Clusters that had significant CBF and miRNA and CBF and VR score correlations were used as input for the mediation analysis, which tests whether the relationship between the independent variable (IV) and dependent variable (DV) is influenced by the mediator (M). Standard mediation analysis is a stepwise process: 1) determine the relationship between IV and DV, 2) determine the relationship between IV and M, 3) determine the relationship between M and DV, 4) determine the relationship between IV and DV controlling for M. If M indeed mediates the relationship between IV and DV, step 4 should be non-significant. To test the directed mediation, the miRNA level was entered as the IV, with either the cluster averaged CBF value (mean cluster) or the CBF value extracted from the voxel with maximum T value (Maxvox) as the mediator, and VR Balance score as the DV. To confirm the mediation effects, we also tested a control mediation where the cluster CBF value was entered as the IV and the miRNA level as the mediator. Sobel's test was used to test for the significance of the mediation. For a mediation to be considered significant, it must satisfy the following criteria: 1) Both Maxvox and mean cluster must have > 70% direct mediation, Sobel's test p-value < 0.1, and 2) Both Maxvox and mean cluster must have < 30% control mediation.

**Results**

Analysis involved 5 steps, with graph evaluation and Cook's Distance analyses – a gold standard engineering approach – performed after inductive statistics to assess effects of potential outlier data [61-63]. We first performed analyses of two-way interactions (1) between rCBF and VR scores, and (2) between rCBF and miRNA levels, followed by analysis of (3) three-way interactions between rCBF, VR, and miRNA to quantify potential overlap between the results of (1) and (2). The three-way analyses were done in two ways – (i) using the rCBF-VR results as a mask for rCBF-miRNA data, and (ii) using the rCBF-miRNA results as a mask for rCBF-VR data to assess consistency of results. We then identified (4) miRNAs showing nominal effects in two-way association with VR measures, and performed (5) directed mediation analysis between rCBF, miRNA and VR measures. For reporting, directed mediation analyses were required to show Sobel p-values < 0.1 across both cluster and voxel with maximum T statistics (maxvox) signals, with percent mediation above 70% in both. Further, control mediation analyses (where miRNA was the mediator and rCBF the independent variable) were required to be non-significant (all Sobel p > 0.1 and mediation effects < 30%) for both maxvox and cluster data.

rCBF and VR scores correlation

Two-way analysis between rCBF and VR scores revealed negative correlations between rCBF and two of the four VR measures: comprehensive score and balance score. A negative correlation means better performance (higher VR score) was associated with lower rCBF. Although the balance score contributed to the comprehensive score, the clusters detected for these two values were in different brain regions, suggesting that the comprehensive score results were not solely driven by the balance scores. Comprehensive score was negatively correlated with rCBF in the thalamus, post cingulate/precuneus, and lateral temporal areas. The balance score, on the other hand, was correlated with rCBF in several regions including the bilateral putamen and in fronto-orbital cortex (FOC). A detailed summary of the location and size of clusters is shown in Table 1. The clusters, overlaid onto a single subject's anatomical image in MNI space, and corresponding scatterplots are shown in Supplemental Figure 1a) results for comprehensive score, and 1b) results for balance score.

**Table 1. Summary of clusters with significant correlations between CBF and VR scores.**

<table>
<tr><th></th><th>Dir</th><th>Nvoxels</th><th>Peak T</th><th>P (unc)</th><th>x,y,z {mm}</th><th>Region</th><th>Label</th></tr>
<tr><td rowspan="6">Comprehensive</td><td>+ve</td><td colspan="6">n.s.</td></tr>
<tr><td rowspan="5">-ve</td><td>285</td><td>8.16</td><td>6.23E-08</td><td>54, -33, 34</td><td>parietal</td><td>R POC (37%), R PT (18%), R anterior SMG (8%), R anterior SMG (8%), R posterior SMG (2%)</td></tr>
<tr><td>299</td><td>7.46</td><td>2.31E-07</td><td>-8, -41, 34</td><td>limbic</td><td>L posterior cingulate (44%), L precuneus (39%)</td></tr>
<tr><td>265</td><td>5.53</td><td>1.22E-05</td><td>3, -2, 4</td><td>sub-lobar</td><td>L Thalamus (53%), R Thalamus (25%)</td></tr>
<tr><td>206</td><td>5.05</td><td>3.54E-05</td><td>45, -39, 5</td><td>temporal</td><td>R posterior SMG (35%), R tempero-occipital MTG (12%), R posterior STG (3%)</td></tr>
<tr><td>210</td><td>4.82</td><td>5.97E-05</td><td>-47, -62, 33</td><td>parietal</td><td>L superior LOC (83%), L angular gyrus (18%)</td></tr>
<tr><td rowspan="2">Spatial Memory</td><td>+ve</td><td colspan="6" rowspan="2">n.s.</td></tr>
<tr><td>-ve</td></tr>
<tr><td rowspan="5">Balance</td><td>+ve</td><td colspan="6">n.s.</td></tr>
<tr><td rowspan="4">-ve</td><td>999</td><td>6.93</td><td>6.59E-07</td><td>-18, 30, -20</td><td>frontal</td><td>L FOC (23%), L subcallosal (9%), L Putamen (20%), L subcallosal (9%), L FMC (4%), L Frontal Pole (2%), L insula (2%)</td></tr>
<tr><td>1219</td><td>5.93</td><td>5.25E-06</td><td>27, 6, -8</td><td>sub-lobar</td><td>R Putamen (34%), R FOC (11%), R subcallosal (9%), R Caudate (8%), R Amygdala (5%), R Pallidum (3%), R Accumbens (3%), R FMC (3%), R Frontal Pole (2%)</td></tr>
<tr><td>388</td><td>5.78</td><td>7.24E-06</td><td>66, -21, -1</td><td>temporal</td><td>R posterior STG (75%), R anterior STG (7%), R posterior MTG (4%), R anterior MTG (2%)</td></tr>
<tr><td>217</td><td>5.15</td><td>2.84E-05</td><td>-5, 9, -11</td><td>frontal</td><td>L Caudate (35%), L Accumbens (34%)</td></tr>
<tr><td rowspan="2">Reaction Time</td><td>+ve</td><td colspan="6" rowspan="2">n.s.</td></tr>
<tr><td>-ve</td></tr>
</table>

The peak T value within the cluster and its location are included in this table. Percentage values in the label column represent the percentage of anatomical region defined on the Harvard-Oxford atlas occupied by the cluster.

Abbreviations for all tables: Dir=direction, +ve=positive, -ve=negative, Nvoxels=number of voxels in cluster, R=right, L=left, POC= Parietal Operculum Cortex, PT= Planum Temporale, SMG= Supramarginal Gyrus, MTG= Middle Temporal Gyrus, STG= Superior Temporal Gyrus, LOC= Lateral Occipital Cortex, FOC= Frontal Orbital Cortex, FMC= Frontal Medial Cortex, COG= Central Opercular Cortex.

**Table 2. Clusters with significant correlations between preseason CBF and miRNA levels.**

| miRNA | Nvoxels | Peak T | P (unc) | x,y,z {mm} | Region | Label |
|---|---|---|---|---|---|---|
| miR-20a (k=208) | 434 | 6.62 | 2.17E-06 | 27, -15, 5 | sub-lobar | R Putamen (46%), R Pallidum (46%) |
| | 230 | 6.62 | 2.19E-06 | 0, -30, -12 | midbrain | Brainstem (90%), R cerebellum I-IV (9%) |
| | 855 | 5.93 | 8.27E-06 | -18, 23, -20 | frontal | L FOC (30%), L Subcallosal (13%), L Putamen (6%), L Paracingulate (4%) |
| | 316 | 5.80 | 1.07E-05 | -17, 2, 9 | sub-lobar | L Thalamus (35%), L Caudate (17%), L Pallidum (4%) |
| | 645 | 5.74 | 1.21E-05 | 27, 20, 3 | sub-lobar | R FOC (27%), R Putamen (18%), R Frontal Pole (10%), R Subcallosal (9%), R Accumbens (3%), R Caudate (2%) |
| | 214 | 5.59 | 1.63E-05 | -48, -17, 21 | parietal | L COG (50%) |
| miR-30d (k=199) | 324 | 7.2 | 5.16E-07 | -41, -6, 18 | sub-lobar | L COG (84%), L Insula (9%), L frontal operculum. (4%), L pars opercularis (1%) |
| | 1752 | 7.2 | 5.54E-07 | 27, 11, -5 | sub-lobar | R Putamen (32%), R Subcallosal (4%), R COG (15%), R FOC (7%), R Pallidum (5%), R Subcallosal (4%), R Insula (3%), R Caudate (3%), R Accumbens (2%), R frontal operculum (2%) |
| | 364 | 5.8 | 8.44E-06 | -23, 17, -1 | sub-lobar | L Putamen (42%), L FOC (19%) |
| | 230 | 5.2 | 2.74E-05 | -18, -9, 15 | sub-lobar | L Thalamus (84%) |
| miR-92a (k=208) | 258 | 5.5 | 1.44E-05 | 27, -15, 5 | sub-lobar | R Putamen (48%), R Pallidum (19%) |
| | 325 | 5.5 | 1.56E-05 | 33, 12, 15 | sub-lobar | R Putamen (33%), R COG (11%), R Insula (10%), R frontal operculum (4%) |
| | 287 | 5.1 | 4.09E-05 | -18, 23, -21 | frontal | L FOC (41%), L Putamen (27%) |
| | 221 | -7.9 | 0.000566 | -21,-55,-18 | post. Cerebellum | L cerebellum V (48%), L cerebellum VI (44%), L lingual (5%), L tempero-occipital Fusiform (2%) |
| miR-195 (k=204) | 246 | 6.4 | 2.64E-06 | -62,-13,15 | parietal | L COG (61%), L Postcentral (16%), L POC (5%), L Heschls (5%) |
| | 244 | -8.62 | 4.19E-08 | 0, 48, -15 | frontal | R FMC (41%), L FMC (38%), R Frontal Pole (16%), L Frontal Pole (6%) |
| | 219 | -5.77 | 8.99E-06 | 12, -57, 6 | limbic | R Lingual (58%), R Precuneus (37%), R Intracalcarine (6%) |
| miR-151-5p (k=193) | 316 | 7.1 | 6.54E-07 | 15, -1, 13 | sub-lobar | L Thalamus (32%), R Thalamus (15%), R Caudate (7%), R Caudate (7%) |

Negative correlations are denoted by a minus sign before the peak T value. See Table 1 for abbreviations.

rCBF and miRNA correlation

Of the nine miRNAs tested for two-way relationship with rCBF, two showed no significant correlations: miR-362-3p and miR-93p. Examination of the correlation plots between rCBF and miRNA levels revealed potential outliers for miR-505* and miR-486, which was confirmed by Cook's Distance analysis. Therefore, results for these two miRNAs are in Supplementary Table 1 and excluded from further analysis. Of the remaining miRNAs that had significant correlations, miR-20a, miR-30d and miR-151-5p had positive correlations, where higher rCBF was associated with higher miRNA levels. For miR-20a, the significant clusters were located in multiple brain regions, including the brainstem and right cerebellum, FOC, caudate, putamen, and thalamus. Similar clusters were found for miR-30d. MiR-151-5p results were limited to the caudate and thalamus. The remaining miRNAs miR-92a and miR-195, showed both positive and negative correlations with rCBF. Interestingly, positive correlations between rCBF and several miRNAs, specifically miR-20a, miR-92a and miR-30d, co-localized to the same clusters involving regions in the basal ganglia and FOC. Table 2 and Supplemental Figure 2 summarize the location and size of these clusters.

rCBF, VR scores and miRNA interaction

Analyses using (a) rCBF-VR balance score results as a mask for the rCBF-miRNA correlations largely overlapped those using the (b) rCBF-miRNA correlations as a mask for the rCBF-VR balance score correlations. Therefore, only the results for (a) are tabulated in Table 3 and shown in Figure 1.

Three of the seven tested miRNAs showed three way relationships with preseason rCBF and VR balance score: miR-20a, miR-30d, and miR-92a. The three way relationships for all three miRNAs were co-localized to two clusters: (1) extending from the left posterior FOC to the left putamen (Put) and insula (INS), and (2) in the right Put. Summary data for these clusters is listed in Table 3 and their corresponding locations in Montreal Neurological Institute (MNI) space and scatterplots are shown in Figure 1. Visualization-based localization, using published neuroanatomic landmarks [64] [65, 66] was performed by a motor control and basal ganglia expert (AJB) to check and complement the automated localization for these clusters, and determined that the left rCBF focus extended from anterior Put to anterior INS/claustrum to sub-callosal cingulate/ventromedial prefrontal cortex and posterior FOC (with some variability across the three miRNAs), whereas the right rCBF focus was localized primarily within the anterior Put, with extension to caudate and/or nucleus accumbens in some cases.

No significant clusters were detected when the rCBF-VR Comprehensive score results were used as a mask for the rCBF-miRNA correlations.

**Table 3. Summary of clusters with significant three-way interactions between preseason CBF, miRNA levels and VR balance score.**

| miRNA | cluster p(FWE) | Nvoxels | Peak T | x,y,z {mm} | Region | Label |
|---|---|---|---|---|---|---|
| miR-20a | 0.002 | 223 | 5.55 | -18, 24, -20 | frontal | L FOC (36%), L Putamen (11%), L subcallosal (3%) |
| | 0.018 | 122 | 4.05 | 21, 14, -11 | sub-lobar | R Putamen (77%), R Accumbens (7%), R Caudate (6%) |
| miR-30d | 6.75E-06 | 503 | 7.18 | 27, 11, -5 | sub-lobar | R Putamen (45%), R FOC (18%), R Caudate (6%), R subcallosal (6%), R Accumbens (3%) |
| | 0.0004 | 287 | 5.80 | -23, 17, 0 | sub-lobar | L Putamen (40%), L FOC (15%) |
| miR-92a | 0.043 | 93 | 4.41 | 27, 11, -5 | sub-lobar | R Putamen (85%) |
| | 0.004 | 184 | 4.37 | -32, 5, 1 | sub-lobar | L Putamen (35%), L FOC (21%) |
| miR-195 | n.s. | | | | | |
| miR-151-5p | | | | | | |

See Table 1 for abbreviations.

**Table 4. Associations between Virtual Reality scores and miRNA levels.**

| Behavioral measure | Statistics | miRNA | | | | | | | | |
|---|---|---|---|---|---|---|---|---|---|---|
| | | 20a | 505 | 362_3p | 30d | 92a | 486 | 195 | 9_3p | 151_5p |
| Comprehensive | Pearson *r* | **-.538** | **-.471** | -.079 | **-.506** | **-.444** | **-.445** | *-.406* | -.332 | **-.544** |
| | p value | **.014** | **.031** | .728 | **.016** | **.039** | **.038** | *.061* | .131 | **.009** |
| | n | 20 | 21 | 22 | 22 | 22 | 22 | 22 | 22 | 22 |
| Balance | Pearson *r* | **-.460** | **-.693** | -.084 | **-.581** | **-.463** | **-.503** | *-.396* | -.230 | **-.634** |
| | p value | **.041** | **.000**** | .711 | **.005**** | **.030** | **.017** | *.068* | .303 | **.002**** |
| | n | 20 | 21 | 22 | 22 | 22 | 22 | 22 | 22 | 22 |

Footnote: Bold indicates a nominally significant association at the p<.05 level, and italics indicate a trend effect. ** indicates results were significant with a Bonferroni correction (i.e. p=0.05/9 miRNAs=0.006); Comprehensive results are presented to guide future work given the three-way analyses showed no significant effects for miRNA-CBF-VR Comprehensive associations.

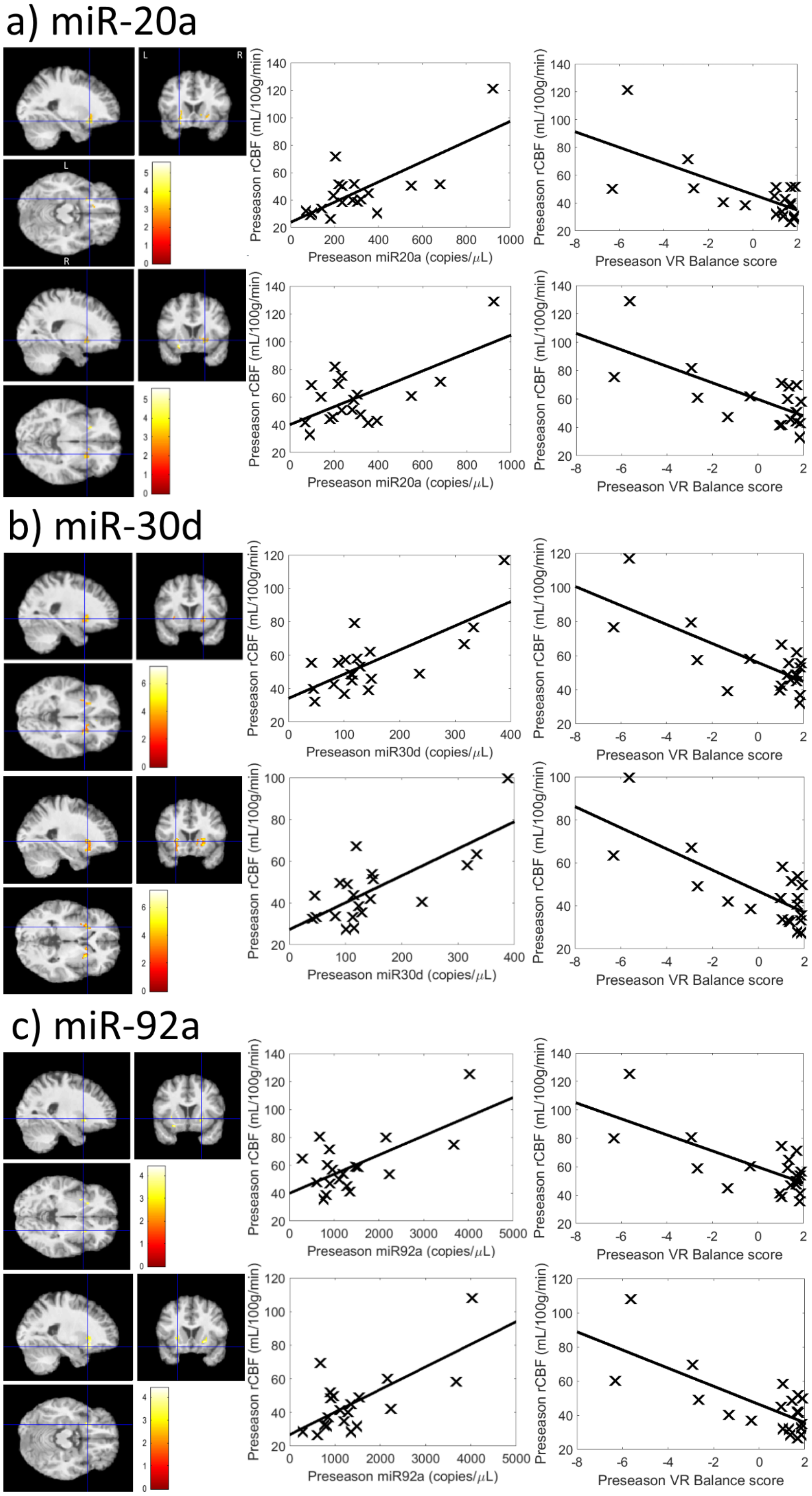

Figure 1. Clusters where significant correlations between CBF and miRNA levels and virtual reality balance scores were detected and their corresponding scatterplots. a) miR-20a results; b) miR-30d results; and c) miR-92a results. Results for miR-505* and miR-486 were excluded as they did not survive Cook's Distance analysis.

miRNA and VR scores correlation

Table 4 shows the Pearson correlation r and p-values for pairwise correlations between preseason miRNA levels and VR scores. Only VR Comprehensive and Balance scores are shown as these were the only scores significantly correlated with CBF in the image-based correlation analysis. Both scores were correlated ($p< 0.05$) with levels of miR-20a, miR-505, miR-30d, miR-92a, miR-486 and miR-151-5p. Additionally, a trend effect was observed for miR-195 levels and both VR Comprehensive and Balance scores ($p\sim0.07$). All of these correlations were negative, where higher levels of miRNA are associated with worse VR performance.

Mediation results

Mediation analysis focused on the two rCBF foci (left Put-INS-FOC and right Put) identified for three miRNAs (miR-20a, miR-30d, miR-92a) from the analysis of interaction between rCBF, VR balance scores, and miRNA. rCBF values extracted from both the maxvox and entire cluster were used for 1) direct mediation, (i.e., miRNA predicting rCBF and rCBF predicting balance), and 2) control mediation, (i.e., rCBF predicting miRNA, miRNA predicting balance). All results are shown in Table 5. Given the number of statistical tests run for the mediation analyses, stringent criteria were used to determine the significance of the mediation analysis results: 1) Directed mediation needed to show a Sobel p-value < 0.1 for both the maxvox and cluster data when rCBF was the mediator variable, with % mediation effects > 70%. 2) Control mediation (rCBF predicting miRNA, miRNA predicting balance) had to be non-significant (Sobel's test $p > 0.1$) with % mediation effects for both maxvox and cluster data being less than 30%. From these steps, conjunction of significant values was below the Bonferroni-corrected threshold for the total number of comparisons made.

Based on the above criteria, the mediation analyses revealed three general outcomes. *First*, two significant results were observed for (a) the right Put, miR-92a and balance score, and (b) the left Put-INS-FOC, miR-30d, and balance score. The average percent mediation effect across both maxvox and cluster analyses was 88% and 89%, respectively, compared to an average of 6% and 8% for the control mediations. These results are shown in **bold** in Table 5. Scatterplots for each of the pairwise correlations relevant for these results, as well as the location of the clusters overlaid onto a high-resolution anatomical image in MNI template space, are shown in Figure 2. Notably, the left Put-INS-FOC maxvox for miR-30d was the strongest result, with a p-value of $p=0.002$, meeting Bonferroni correction for all Sobel tests computed (12 directed mediation and 12 control mediations or $p < 0.05/24 = .0021$). The maxvox rCBF in this case mediates 100% of the relationship between miR-30d levels and VR balance, while the control mediation was 0%. *Second*, trend effects were noted for (c) the right Put, miR-20a, and balance score, and (d) the left Put-INS-FOC, miR-92a, and balance score. These results were considered trend effects as only the mean cluster rCBF satisfied the criteria for significance described above. They are denoted by * in Table 5. *Third*, non-significant or inconsistent effects were noted for (e) the right Put, miR-30d, and balance score, and (f) the left Put-INS-FOC, miR-20a, and balance score, where either none of the direct mediation results were significant (f), or the percent mediation was higher in the control mediation case (e). A detailed description of the most robust findings (i.e. (a) and (b)) is included below.

*miR-30d, left Put-INS-FOC, and VR balance* Pairwise regressions between (1) miR-30d and cluster rCBF in the left Put-INS-FOC ($\beta=0.737$, $p<0.001$), (2) cluster rCBF and VR Balance score ($\beta=-0.740$, $p=0.001$) and (3) miR-30d and VR Balance score ($\beta=-0.614$, $p=0.004$) were all significant ignoring the mediator (cluster rCBF). After accounting for the effect of the mediator, the relationship between miR-30d and VR Balance score was no longer significant ($\beta=-0.151$, $p>0.2$), with the mediator explaining 75%, and Sobel $p=0.022$. Control mediation analysis, wherein miRNA was the mediator, showed 15% effect mediated, and Sobel $p>0.2$.

For this three-way relationship and mediation, there was also a salient result for the maxvox of the left Put-INS-FOC. Specifically, pairwise regressions between (1) miR-30d and maxvox rCBF in the left Put-INS-FOC ($\beta=0.807$, $p<0.001$), (2) maxvox rCBF and VR Balance score ($\beta=-0.807$, $p=0.001$) and (3) miR-30d and VR Balance score ($\beta=-0.614$, $p=0.004$) were all significant ignoring the mediator (maxvox rCBF). After accounting for the effect of the mediator, the relationship between miR-30d and VR Balance score was no longer significant ($\beta=0.106$, $p>0.2$), with the mediator explaining 100%, and Sobel $p=0.002$ (meeting a Bonferroni correction for all mediation analyses run). Control mediation analysis, wherein miRNA was the mediator, showed 0% effect mediated, and Sobel p>0.2.

*miR-92a, right Put, and VR balance* Pairwise regressions between (1) miR-92a and cluster rCBF in the right Put (β=0.691, p<0.001), (2) cluster rCBF and VR Balance score (β=-0.722, p=0.001) and (3) miR-92a and VR Balance score (β=-0.465, p=0.039) were all significant ignoring the mediator (cluster rCBF). After accounting for the effect of the mediator, the relationship between miR-30d and VR Balance score was no longer significant (β=0.066, p>0.2), with the mediator explaining 100%, and Sobel p=0.010. Control mediation analysis, wherein miRNA was the mediator, showed 0% effect mediated, and Sobel p>0.2.

For this three-way relationship and mediation, there was also a trend effect for the maxvox of right Put. Specifically, pairwise regressions between (1) miR-92a and maxvox rCBF in the right Put (β=0.720, p<0.001), (2) maxvox rCBF and VR Balance score (β=-0.589, p=0.006) and (3) miR-92a and VR Balance score (β=-0.465, p=0.039) were all significant ignoring the mediator (maxvox rCBF). After accounting for the effect of the mediator, the relationship between miR-92a and VR Balance score was no longer significant (β=-0.085, p>0.2), with the mediator explaining 82%, and Sobel p=0.085. Control mediation analysis, wherein miRNA was the mediator, showed 10% effect mediated, and Sobel p>0.2.

**Table 5. Mediation Results**

| **Model 1**: IV Path A → M Path B → DV<br>**Model 2**: IV Path C → DV | | | | **Path A: IV predicting Mediator** | | **Path B: Mediator Predicting DV** | | **Path C: IV predicting DV (model 1)** | | **Path C: IV (with mediator) predicting DV (model 2)** | | **Effect Mediated** | **Sobel Test** |
|---|---|---|---|---|---|---|---|---|---|---|---|---|---|
| | | **IV** | **M** | **Std β** | **p** | **Std β** | **p** | **Std β** | **p** | **Std β** | **p** | **%** | **p** |
| R Put (21, 14, -11) | Directed | miR-20a | MaxVox | .701 | .001 | -.571 | .011 | -.469 | .043 | -.135 | .644 | 71 | .123 |
| | | miR-20a | Mean Cluster | .643 | .003 | -.700 | <.001 | -.469 | .043 | -.031 | .894 | 93* | .026* |
| | Control | MaxVox | miR-20a | .701 | <.001 | -.469 | .043 | -.571 | .011 | -.476 | .115 | 17 | >0.20 |
| | | Mean Cluster | miR-20a | .643 | .003 | -.469 | .043 | -.700 | <.001 | -.679 | .010 | 3* | >0.20* |
| L Put-INS-FOC (-18, 24, -20) | Directed | miR-20a | MaxVox | .803 | <.001 | -.553 | .014 | -.469 | .043 | -.071 | .842 | 85 | .169 |
| | | miR-20a | Mean Cluster | .736 | <.001 | -.685 | .001 | -.469 | .043 | .078 | .775 | 85 | .169 |
| | Control | MaxVox | miR-20a | .803 | <.001 | -.469 | .043 | -.553 | .014 | -.496 | .174 | 10 | >0.20 |
| | | Mean Cluster | miR-20a | .736 | <.001 | -.469 | .043 | -.685 | .001 | -.743 | .014 | 0 | >0.20 |
| R Put (27,11,-5) | Directed | miR-30d | MaxVox | .861 | <.001 | -.589 | .006 | -.614 | .004 | -.416 | .279 | 32 | >0.20 |
| | | miR-30d | Mean Cluster | .737 | <.001 | -.744 | <.001 | -.614 | .004 | -.145 | .550 | 76 | .020 |
| | Control | MaxVox | miR-30d | .861 | <.001 | -.614 | .004 | -.589 | .006 | -.230 | .544 | 61 | >0.20 |
| | | Mean Cluster | miR-30d | .737 | <.001 | -.614 | .004 | -.744 | <.001 | -.638 | .015 | 14 | >0.20 |
| **L Put-INS-FOC (-23,17, 0)** | **Directed** | **miR-30d** | **MaxVox** | **.807** | **<.001** | **-.807** | **<.001** | **-.614** | **.004** | **.106** | **.666** | **100** | **.002** |
| | | **miR-30d** | **Mean Cluster** | **.737** | **<.001** | **-.740** | **<.001** | **-.614** | **.004** | **-.151** | **.536** | **75** | **.022** |
| | **Control** | **MaxVox** | **miR-30d** | **.807** | **<.001** | **-.614** | **.004** | **-.807** | **.000** | **-.893** | **.002** | **0** | **>0.20** |
| | | **Mean Cluster** | **miR-30d** | **.737** | **<.001** | **-.614** | **.004** | **-.740** | **<.001** | **-.629** | **.017** | **15** | **>0.20** |
| **R Put (27,11,-5)** | **Directed** | **miR-92a** | **MaxVox** | **.720** | **<.001** | **-.589** | **.006** | **-.465** | **.039** | **-.085** | **.768** | **82** | **.085** |
| | | **miR-92a** | **Mean Cluster** | **.647** | **.002** | **-.702** | **<.001** | **-.465** | **.039** | **-.018** | **.938** | **96** | **.020** |
| | **Control** | **MaxVox** | **miR-92a** | **.720** | **<.001** | **-.465** | **.039** | **-.589** | **.006** | **-.528** | **.079** | **10** | **>0.20** |
| | | **Mean Cluster** | **miR-92a** | **.647** | **.002** | **-.047** | **.039** | **-.702** | **<.001** | **-.691** | **.007** | **2** | **>0.20** |
| L Put-INS-FOC (-32,5,1) | Directed | miR-92a | MaxVox | .717 | <.001 | -.526 | .017 | -.465 | .039 | -.180 | .546 | 61 | .196 |
| | | miR-92a | Mean Cluster | .691 | <.001 | -.722 | <.001 | -.465 | .039 | .066 | .780 | 100* | .010* |
| | Control | MaxVox | miR-92a | .717 | <.001 | -.465 | .039 | -.526 | .017 | -.396 | .194 | 25 | >0.20 |
| | | Mean Cluster | miR-92a | .691 | <.001 | -.047 | .039 | -.722 | <.001 | -.768 | .004 | 0* | >0.20* |

rCBF from either the voxel with the maximum T-value (maxvox) or mean over the entire cluster (mean cluster) was used in these analyses. Results were considered significant if a) both maxvox and mean cluster show above 70% mediation and have a Sobel's test p-value < 0.1 and b) the control mediation is less than 30% and has Sobel's test p-value > 0.1. Both the Left Put-INS-FOC cluster of miR-30d and right Put cluster of miR-92a satisfy these criteria and are shown in bold. Results where only maxvox or mean cluster but not both satisfy the above criteria are marked with *, these include the right Put cluster for miR-20a, R Put-INS-FOC cluster for miR-30d and Left Put-INS-FOC cluster for miR-92a.

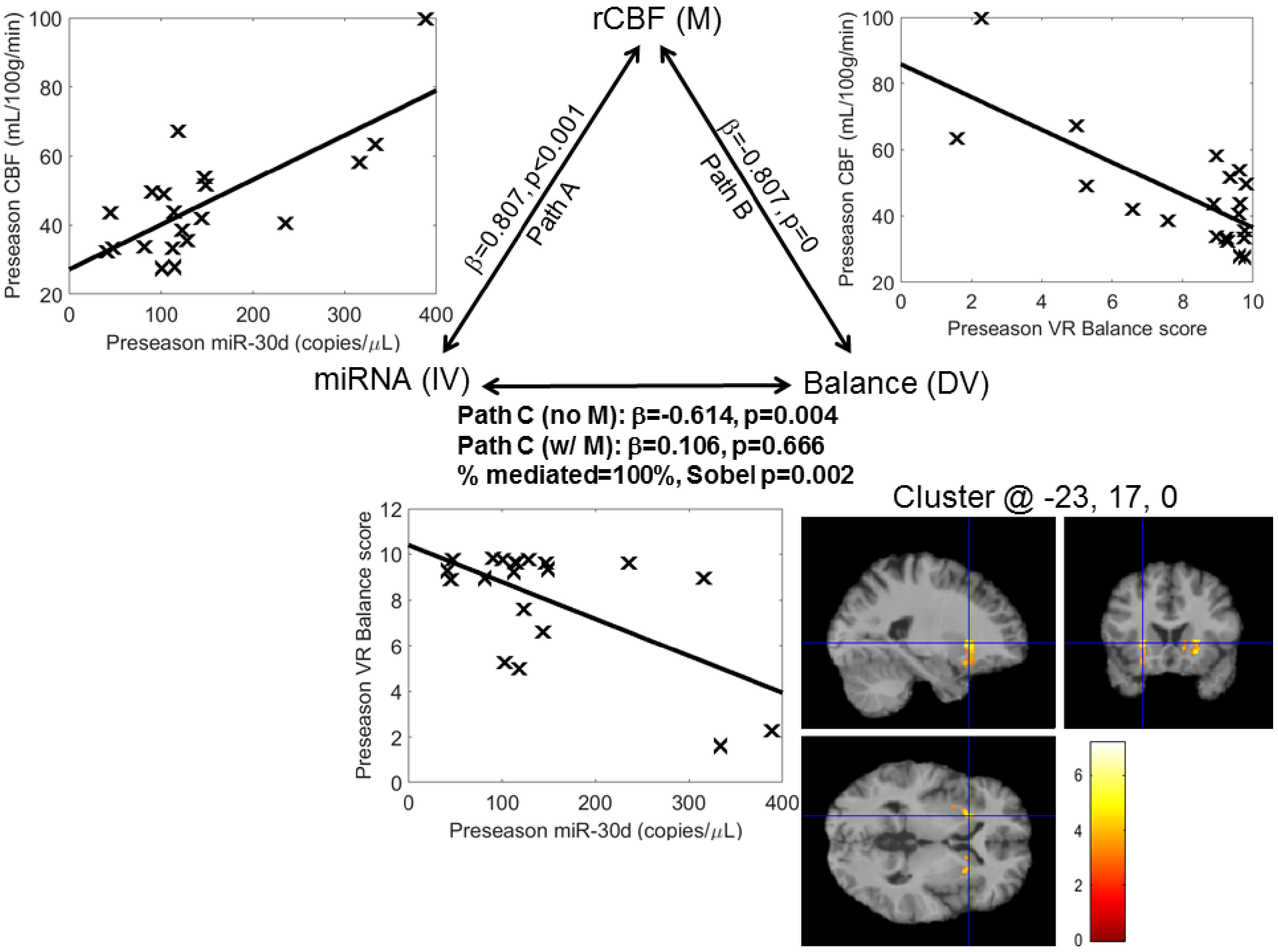

Figure 2. Mediation results for miR-30d, extracted from the L Put-INS-OFC cluster.

## Discussion

This study examined whether a previously identified panel of miRNAs distinctly elevated among football players exposed to HAEs [25] is associated with rCBF and postural control in a cohort of NCAA Division I football athletes, and if rCBF mediated the effects of miRNA on behavior. We hypothesized that elevated miRNA levels observed before seasonal play for collegiate football players were potentially residual signs of inflammation as supported by the presence of a three-way interaction between pre-season miRNA levels, rCBF of brain regions producing motor control behavior, and postural control measured by sensitive VR tasks. We asked if perfusion in basal ganglia, SMA, and cerebellar regions connected to these motor control functions showed a significant association with these behaviors, potentially as a negative association given other published data [35]. We further asked if rCBF in these regions were related to baseline miRNA levels, potentially in a positive fashion as might be expected if elevated inflammation led to increased perfusion. We further sought a directed mediation relationship between these variables that was consistent across mediation relationships, and posited rCBF as the mediating variable between miRNA and behavior, in a manner analogous to the idea of an intermediate phenotype in imaging genetics [67, 68], yet moving mechanistically beyond this perspective. Together, these observations argued for a model of ongoing residual inflammation in these return athletes.

Our study supports this model with a number of robust findings: (1) rCBF was negatively correlated with VR Balance and Comprehensive scores (i.e., elevated rCBF was associated with worsened VR performance) uniformly in task-relevant regions, such as the left Put-INS-FOC and right Put. (2) rCBF was differentially correlated with a subset of the miRNAs tested, and where there was a three way relationship between miRNA, rCBF, and VR behavior, it was always positively associated with miRNA levels. (3) MiRNA levels and VR behavior showed only negative associations, so that higher miRNA levels were associated with worsening behavior. (4) For the three-way interactions observed, rCBF consistently mediated the detrimental effects of miRNAs on VR Balance, and there was no partial mediation with miRNA mediating the association of rCBF with VR behavior.

Mediation analyses specifically focused on three miRNAs and two basal ganglia loci involving the bilateral Put (Table 5). Two of these three analyses were quite robust – one involving miR-30d and the left Put-INS-FOC, and the other involving miR-92a and the right Put. Given the robust effects mediated (and not mediated by the control analyses), these results should be replicable in other HAE athlete samples. Two other results were suggestive, but will likely require larger samples for retesting as the cluster and maxvox results did not agree – these included miR-20a and the right Put, and miR-92a and the left Put-INS-FOC. Lastly, two of these results showed inconsistent findings between maxvox and mean cluster data (miR-30d and right Put-INS-FOC), or low Sobel p-values for maxvox and mean cluster data (miR-20a and left Put); larger sample sizes are likely to produce different results from those reported herein. It should be noted, that the directionality of the three-way relationships observed was consistent across all 6 analyses, and support the following general model: (i) chronic HAEs induce a persistent neuroinflammatory response; (ii) the putamen, thought to be fundamental in motor control processes such as balance/postural stability [69-76], is affected by HAEs and chronic inflammation; (iii) chronic putaminal inflammation is associated with impaired behavior. In the text that follows, four general topics are discussed in more detail: (i) the relationship of motor control behavior to rCBF loci, (ii) relationship of miRNA to neuroinflammation, (iii) the relevance of mediation as opposed to intermediary phenotype relationships, and (iv) the assessment of confounds.

*Better VR performance is associated with lower rCBF in task-relevant brain regions*

Negative correlations were found between rCBF and VR Comprehensive and Balance scores in all our analyses. For the VR Comprehensive score, rCBF from brain regions such as the thalamus, posterior cingulate and bilateral SMG were negatively associated with VR score. The posterior cingulate and the adjacent retrosplenial cortex, have long been found to be associated with spatial orientation and working memory, as is evident from both animal and human lesion models [77-79]. Primate studies have shown that these regions receive afferents from the anterior thalamus [80], an area also revealed to be significantly correlated to VR Comprehensive score in the current study. The correlations between rCBF in these regions and VR Comprehensive score are likely driven by the spatial memory and orientation (Reaction Time) components of the VR task. On the other hand, the VR Balance score was associated with brain perfusion magnitude in the putamen, which is thought to be involved in postural and balance control, in addition to other aspects of motor function [69-76]. It is particularly interesting that the Balance score was not associated with perfusion in the cerebellum, given its more established role in balance function relative to the basal ganglia. [69] and [70] hypothesized that the basal ganglia, and particularly the putamen, coordinate(s) different aspects of balance and postural control across many different states and contexts by coordinating recruitment of programs across multiple motor regions (including the cerebellum). It may be that adaptive, dynamic postural responses such as those required by the VR task are linked most tightly to function in the putamen, even if other regions involving balance function (such as the cerebellum) are also impacted by HAEs.  Interestingly, only the VR Balance score was concurrently related to rCBF and miRNA levels, with strong mediation results.

*MicroRNA accumulation is detrimental to VR performance, and may be related to neuroinflammation*

In a previous study on the same cohort, Papa et al. (2018) reported that the preseason miRNA levels in these football athletes were significantly elevated compared to healthy young non-athlete controls [25]. Accumulation of these miRNAs appeared to be detrimental to the athletes' performance on VR tasks, as Papa et al. (2018) also reported that athletes with higher Standardized Assessment of Concussion (SAC) scores (i.e. better neurocognitive performance) had lower miRNA levels. This is corroborated in our correlation analysis, which revealed a negative correlation between miRNA levels and VR performance. Of the three miRNAs that emerged with a significant three-way interaction with rCBF and VR Balance score, miR-20a and miR-92a are part of the miR-17-92 cluster on chromosome 13 [81]. Micro-RNA clusters are comprised of multiple miRNAs that are transcribed by physically adjacent sequences and often target a common factor or signaling pathway. The miR-17-92 cluster has a crucial role in the immune system, as mice deficient in this cluster die shortly after birth [82] and overexpression of this cluster leads to lymphoproliferative disease and autoimmune disorders [83]. MiR-20a, in particular, has been shown to inhibit the secretion of cytokines such as IL-6, IL-8 and IL-10 [84], indicative of its involvement in neuro-inflammation processes. Further evidence of this involvement is seen in a murine model of traumatic spinal cord injury, where miR-20a is upregulated. Infusion of miR-20a induces inflammation and motor neuron degeneration in uninjured spinal cord, while inhibition of miR-20a leads to improved neuronal survival and increasing neurogenesis [85]. Other members of the MiR-17-92 cluster have been associated with inhibition of antiangiogenic genes such as thrombospodin-1 and connective tissue growth factor, suggesting a role in the control of neovascularization [86]. miR-92a, in particular, blocks angiogenesis in both in vitro and in vivo mice models of limb ischemia and myocardial infarction [87]. The third miRNA, miR-30d, has been observed to control apoptosis [88, 89] and promote angiogenesis and tumor growth in an in vivo mouse model [90]. While miRNA studies of brain injury models are still nascent, the aforementioned studies suggest that the elevated miRNA levels observed in the current cohort are related to chronic neuroinflammatory processes, which adversely impact neurocognitive performance.

*rCBF in putamen mediates the relationship between miRNAs and VR Balance*

We identified a three-way interaction between rCBF, miRNA levels and VR Balance in bilateral putamen. Mediation analyses revealed that rCBF in the putamen mediates the relationship between miRNA levels and VR Balance, with percent mediation ranging from 75% to 100%. All of the control mediations, which tested for the mediation effect of miRNA levels on the relationship between rCBF and VR Balance, had percent mediation less than 30%. These findings highlight rCBF as a necessary variable for two separate miRNAs (miR-30d and miR-92a) to affect behavior, and suggest a possible mechanism in place of overlapping associations (i.e., intermediary phenotype). In both hemispheres, clusters were localized to the rostral portion of the putamen, but were relatively more dorsal in the left hemisphere and ventral in the right hemisphere. It is important to note that while the right-hemisphere cluster was primarily centered within the putamen, the left-hemisphere cluster was at the anterior and lateral border of the putamen and extended into white matter and cortical regions known to project more densely to the striosomes than to the matrix compartment of the putamen [91]. Whether this disparity is due to the predominant right-handedness of the present cohort (100% right-handed) or to some functional hemispheric difference within the putamen (e.g., [92]) remains to be elucidated in a larger cohort.

The putamen, together with the caudate, form the striatum and are the input nuclei of the basal ganglia, which receive extensive direct afferents from the cortex. Recent works suggest that cortical inputs to the basal ganglia are organized both topographically and functionally, where the rostral putamen receives inputs relatively more densely from the prefrontal, orbitofrontal and anterior cingulate cortices [93-96], and the caudal putamen receiving inputs mostly from parietal, temporal, pre-supplementary and supplementary motor areas [96-100]. This organization may also relate to the rostral-ventral gradient of striosome versus matrix compartments in these nuclei [91], which may be relevant to identifying the functional networks or systems impacted by HAEs. It is of note that the striosome-dominant cortical domains included in the left putaminal cluster associated with VR Balance and miRNAs (including INS, FOC, and subcallosal cingulate) have been shown to exhibit altered connectivity with the putamen in the genetic disorder, X-linked dystonia parkinsonism [101].

The striatal nuclei also receive input from the cerebellum via a disynaptic connection via the thalamus [102], and project back to the cerebellum also via a disynaptic connection via the subthalamic nucleus[103] . The cerebellum is known to be involved in postural control and balance [71, 104, 105], and may work together with the basal ganglia to coordinate different subcomponents of such functions, in addition to other coordinating other non-motor functions (reviewed in [46]). Although cerebellar perfusion related to levels of two miRNAs (miR-20a and miR-92a) evaluated in this study, this region was not directly correlated with motor or behavioral output measures. However, it is possible that abnormal perfusion in this region contributed in some less direct way to these measures.

The role of the basal ganglia in movement control was first recognized from clinical observations of movement disorders associated with lesions in basal ganglia nuclei [106, 107]. The classic circuitry model of the basal ganglia in motor control includes afferents from the cortex and projections back to the cortex via the output nuclei (globus pallidus interna and substantia nigra pars reticulata) and thalamus [108]. The classic functional model of the basal ganglia proposes that movement is achieved via a delicate balance between direct and indirect pathways between the striatum and pallidi [109, 110]. The putamen is viewed as being involved in basic motor function more than the caudate, which is thought to be involved more in complex behaviors [111-114]. Today, it is widely recognized that there are multiple loops through the basal ganglia, organized topographically and acting in parallel to support multiple motor, limbic, and cognitive functions such as learning, habit formation, executive control and reward processing [115-117]. In addition, there is a growing body of literature evaluating the function of pallidal projections to brainstem nuclei (e.g., [118]; [119-121]). This literature led to the proposal that the extensive network of motor regions connected to the basal ganglia coordinate, in parallel, the selection of movement (direct pathway) and control of whole body stability and posture required both at rest and during movement (indirect pathway) [69, 70].

The athletes in the current study have an average of 11 years of participation in contact sports. The fact that elevated miRNA levels were found at a timepoint when these athletes had at least 16 weeks of rest from HAEs suggests the presence of a chronic neuro-inflammatory process unrelated to acute or recent head impacts. While the miRNA levels caused widespread changes in rCBF within multiple brain regions, only rCBF in bilateral putamen was associated with the impaired VR balance. These results provide preliminary evidence of the utility of an integrative systems neuroscience approach to study abnormal perturbations such as those caused by repetitive HAEs, and may have general applicability in other illnesses. In the future, abnormal miRNAs and VR balance task performance may be used in return to play decision by identifying athletes that need additional recovery and/or treatment. Longer-term application include the development of anti-inflammatory drugs and/or transcriptome or metabolome products that target the putamen to alleviate postural control impairment.

*Study limitations and Assessment of Confounds*

Although the results presented here were robust for miR-30d and miR-92a, some of the other results were not as strong, likely due to the modest cohort size, arguing for replication in a larger cohort. It is also important to note that the current cohort consisted of only male athletes, so we were unable to assess the possibility of gender differences. Future studies with females will be important to generalize these results across both genders.

To assess the possibility of confounds, we incorporated control mediation analyses that were predicted to be non-significant. Namely, we used miRNA levels as the mediator variable to determine if it worsened the relationship between rCBF and Balance scores. In no case did miRNA levels act as a mediator variable, thereby serving as a consistent negative control. This assessment is crucial to understand the specific three way relationships we found in this study as others have argued that mediation analyses which include (a) precedence information (i.e., miRNA is always the independent variable and behavior is the output variable) and (b) confounding controls to rule out non-spuriousness, allow better understanding of potential mechanisms of injury, even in a cross-sectional analyses that does not include longitudinal information or an intervention [58, 59]. Showing that regional brain imaging variables were needed for the effects of miRNA on behavior goes beyond the traditional construct of an intermediary phenotype, where the intermediary phenotype acts as a node in an overlapping set of associations.

*Conclusions*

In this study, we used a novel imaging-omics approach to study the relationship between rCBF, miRNA levels and motor control behavior in NCAA Division I male football athletes. Mediation analysis revealed that rCBF is a significant mediator for the detrimental effects of abnormal inflammatory miRNA levels on motor control. This finding points to the presence of a chronic neuro-inflammatory process in these athletes due to their history of participation in contact sports, and the timing of the study to be after 16 weeks without HAEs. This study offers preliminary evidence for the utility of an integrative systems neuroscience approach to study abnormal perturbations due to processes such as repeated HAEs, and suggests multiple applications to the study of other phenomena affecting the brain.

**Acknowledgements**

All work in this paper was funded by the listed academic institutions, and without specific NIH, NSF, or DoD support. We thank the Penn State football players for their effort participating in this study. We also thank Katie Finelli and Madeleine Scaramuzzo for their assistance with subject recruitment and data collection.

**Author Contributions**

Conceptualization: HCB; Investigation: SMS, PHS, AEW; Methodology: HCB; Software: YFC and SMS; Formal Analysis: YFC, ZM, and JH; Resources: AEW, SMS, AJB, NV, AAH; Data Curation: LP, MB, BKR, AEW; Writing – Original Draft: YFC and HCB; Writing – Review & Editing: all co-authors; Supervision: EN, TMT, LP, SMS, HCB.

**Declaration of interests**

Drs. Papa and Bhomia are inventors of a US patent application filed by Uniformed Services University of the Health Sciences (USUHS) regarding the potential utilities of selected miRNAs as diagnostic biomarkers for TBI. The other authors declare they have no financial or other conflicts of interest with regard to the data and analyses presented herein.

The opinions expressed herein are those of the authors and are not necessarily representative of those from their respective institutions.

## Supplemental Materials

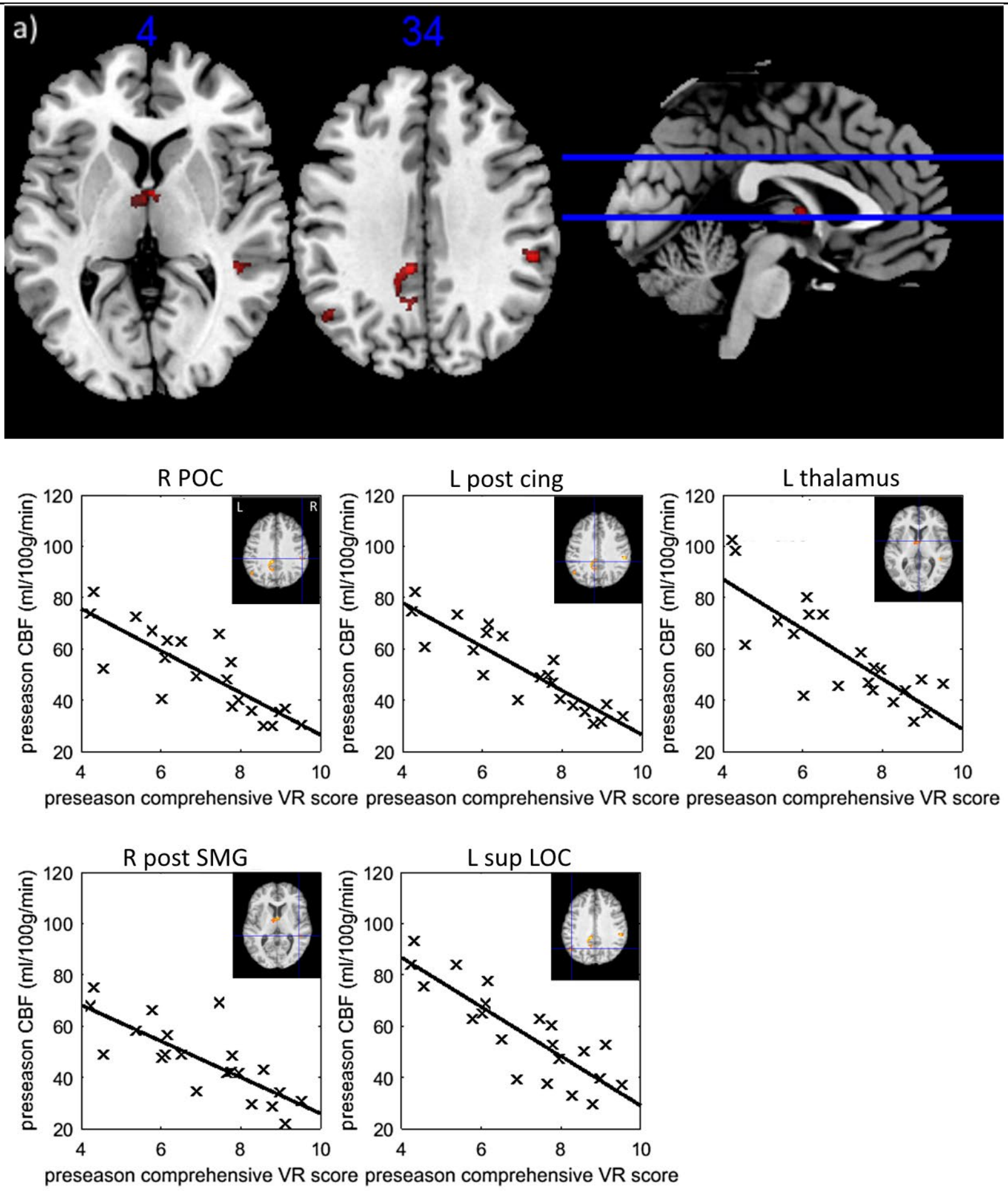

Supplemental Figure 1a) Clusters with significant associations between preseason CBF and preseason comprehensive VR score. Correlation plots extracted from these clusters are shown below.

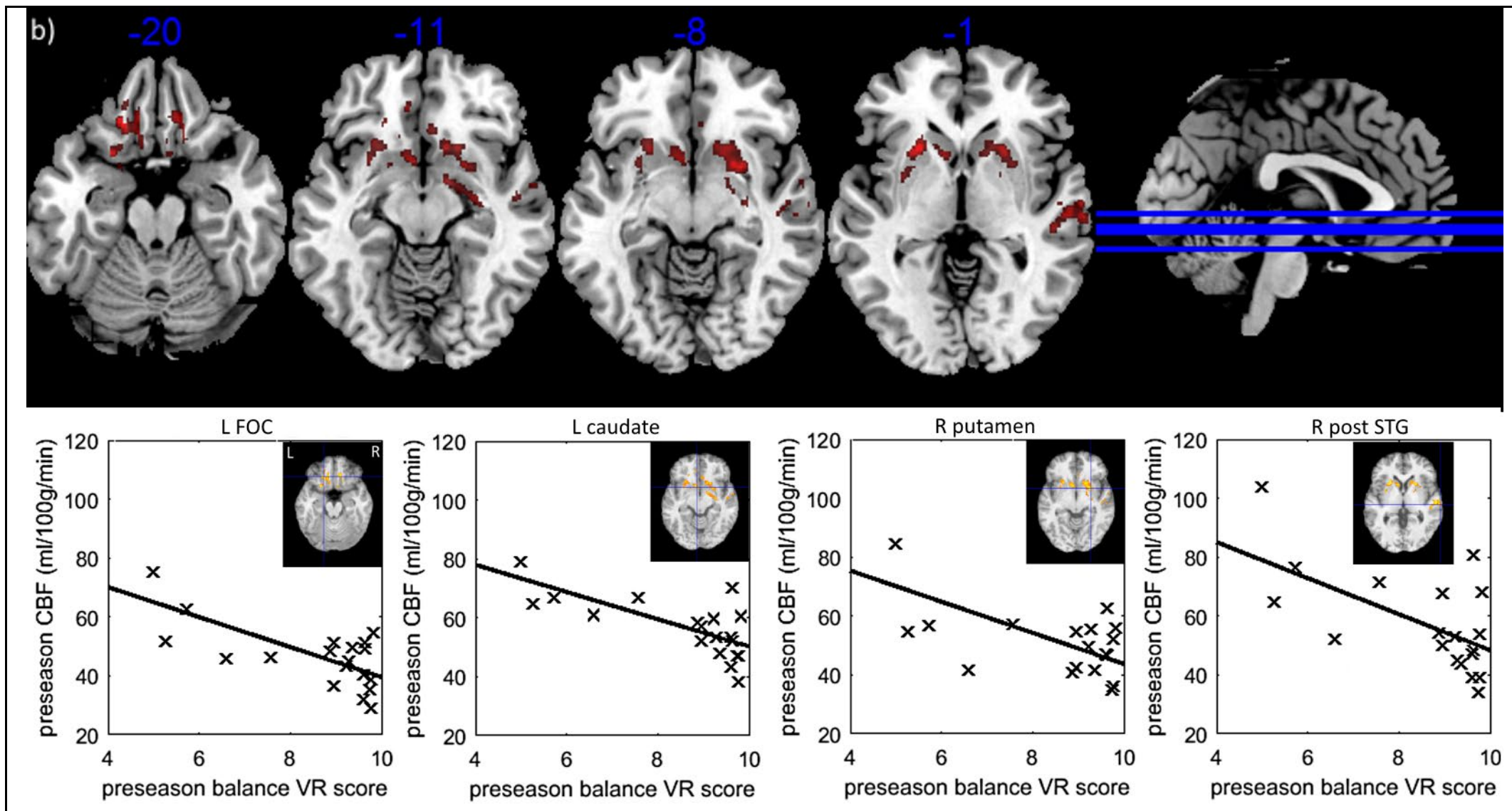

Supplemental Figure 1b) Clusters with significant associations between preseason CBF and preseason balance VR score. Correlation plots extracted from these clusters are shown below.

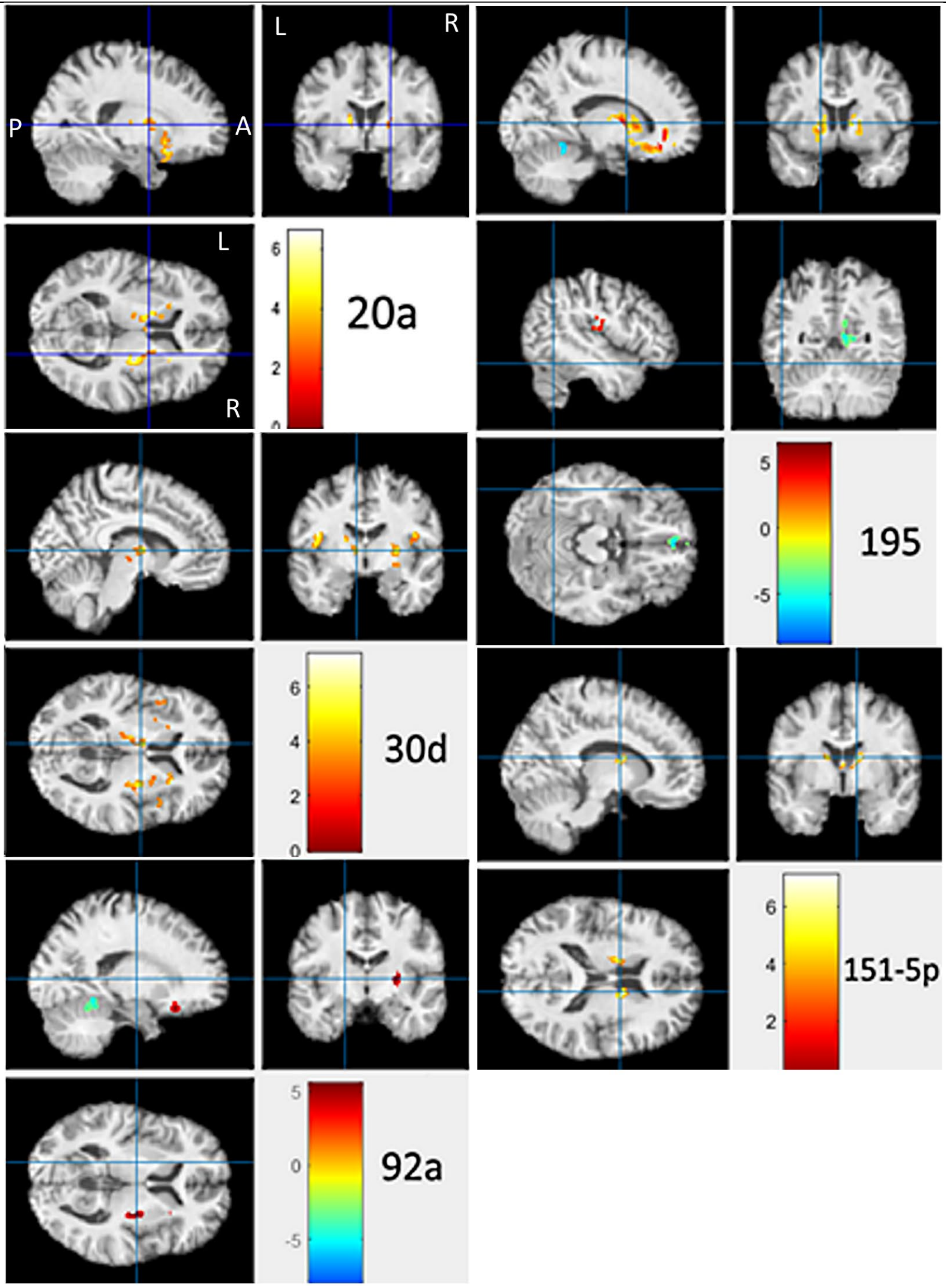

Supplementary Figure 2. Clusters with significant associations between preseason CBF and miRNA levels, overlaid onto a single subject's high-resolution anatomical images.

**Supplemental Table 1. Clusters with significant correlations between preseason CBF and miRNA levels of miR-505* and miR-486.**

| miRNA | dir | Nvoxels | Peak T | p(unc) | x,y,z {mm} | region | label |
|---|---|---|---|---|---|---|---|
| miR-505* (k=200) | +ve | 1674 | 7.4 | 5.12E-07 | 26, -11, 6 | sub-lobar | R. Putamen (27%), R. Pallidum (10%), R. FOC (9%), R. Caudate (5%), R. Subcall (4%), R. POC (3%), R. Thalamus (2%), R. Insula (2%), R. Accumbens (2%) |
| | | 997 | 6.6 | 2.06E-06 | -18, 24, -21 | frontal | L. FOC(19%), L. Subcall (17%), L. Caudate (8%), L. Putamen (4%), L. Accumbens (4%), L. Paracing (2%), L. FMC (2%) |
| | -ve | 278 | 7.6 | 3.91E-07 | -14, -51, -14 | ant. Cerebellum | L. V (57%), L. VI (35%), L. lingual (6%), L. cerebellum I-IV (2%) |
| miR-486 (k=214) | +ve | 2437 | 11 | 1.21E-09 | 26, -17, 2 | sub-lobar | R. Putamen (21%), R. FOC(10%), R. Pallidum (8%), R. Subcall (5%), R. FMC (4%), R. Thalamus (3%), R. Frontal Pole (3%), R. Caudate (3%), R. POC (3%), R. Accumbens (2%), R. Insula (2%) |
| | | 2040 | 9.1 | 1.76E-08 | -11, 26, -11 | limbic | L. FOC(18%), L. Subcall(14%), L. Putamen (10%), L. Caudate (8%), L. Pallidum (3%), L. Thalamus (3%), L. Paracing (2%), L. Accumbens (2%) |
| | | 445 | 7.9 | 1.57E-07 | 0, -30, -10 | midbrain | Brain-Stem (87%) |
| | -ve | 314 | 6.5 | 2.13E-06 | -21, -57, -20 | limbic | L. cerebellum V (52%), L. cerebellum VI (46%), L. Lingual (2%), L. temp. occ. Fusiform (1%) |
| | | 357 | 5.2 | 3.18E-05 | -32, 29, -1 | frontal | L. FOC (82%), L. Frontal Pole (4%), L. Temporal Pole (4%), L. Insula (2%) |

These results did not survive Cook's Distance analysis and were excluded from three-way and mediation analyses. They are included here for reference only.